\documentclass[reprint,superscriptaddress,preprintnumbers,amsmath,amssymb,aps,prl,tightenlines,longbibliography,nofootinbib]{revtex4-2}

\usepackage[T1]{fontenc}
\usepackage[utf8]{inputenc}
\usepackage{lmodern}

\usepackage{graphicx}
\usepackage{mathtools}
\usepackage{physics}
\usepackage{mathrsfs}
\usepackage[shortlabels]{enumitem}
\usepackage[english]{babel}
\usepackage{tikz}
\usetikzlibrary{decorations.pathmorphing, decorations.markings, arrows.meta}

\usepackage[colorlinks=true
    ,urlcolor=blue
    ,anchorcolor=blue
    ,citecolor=blue
    ,filecolor=blue
    ,linkcolor=red
    ,menucolor=blue
    ,linktocpage=true
    ,pdfproducer=medialab
    ,pdfa=true
]{hyperref}

\newcommand{\C}{\mathbb{C}}

\DeclareMathOperator{\sgn}{sgn} 

\DeclareRobustCommand{\smallparallelepiped}{%
  \mathord{\vcenter{\hbox{%
  \tikz[
    x=0.10em,
    y=0.10em,
    line width=0.30pt,
    line cap=round,
    line join=round
  ]{
    \draw (0,0) -- (5.0,0.0) -- (6.0,3.0) -- (1.0,3.0) -- cycle;

    \draw (1.7,1.1) -- (6.7,1.1) -- (7.7,4.1) -- (2.7,4.1) -- cycle;

    \draw (0,0) -- (1.7,1.1);
    \draw (5.0,0.0) -- (6.7,1.1);
    \draw (6.0,3.0) -- (7.7,4.1);
    \draw (1.0,3.0) -- (2.7,4.1);
  }}}}%
}

\makeatletter
\newcounter{savesection}
\newcounter{apdxsection}
\renewcommand\appendix{\par
    \setcounter{savesection}{\value{section}}%
    \setcounter{section}{\value{apdxsection}}%
    \setcounter{subsection}{0}%
\gdef\thesection{\@Alph\c@section}}
\newcommand\unappendix{\par
    \setcounter{apdxsection}{\value{section}}%
    \setcounter{section}{\value{savesection}}%
    \setcounter{subsection}{0}%
\gdef\thesection{\@arabic\c@section}}

\newcommand*\rel@kern[1]{\kern#1\dimexpr\macc@kerna}
\newcommand*\widebar[1]{%
  \begingroup
  \def\mathaccent##1##2{%
    \rel@kern{0.8}%
    \overline{\rel@kern{-0.8}\macc@nucleus\rel@kern{0.2}}%
    \rel@kern{-0.2}%
  }%
  \macc@depth\@ne
  \let\math@bgroup\@empty \let\math@egroup\macc@set@skewchar
  \mathsurround\z@ \frozen@everymath{\mathgroup\macc@group\relax}%
  \macc@set@skewchar\relax
  \let\mathaccentV\macc@nested@a
  \macc@nested@a\relax111{#1}%
  \endgroup
}
\makeatother

\begin{document}

\title{The Positivity Geometry of Photon--Dark-Photon Effective Field Theories}
\date{\today}

\author{Sayantan Chakraborty}
\email{sayantan.chakraborty@students.iiserpune.ac.in}

\author{Yash Dadhwal}
\email{yash.dadhwal@students.iiserpune.ac.in}

\author{Arun M. Thalapillil}
\email{thalapillil@iiserpune.ac.in}

\affiliation{Indian Institute of Science Education and Research (IISER) Pune, Pune 411008, India}

\begin{abstract}
We derive positivity bounds on the complete dimension-eight effective field theory of photons and a massless dark photon. The mixed gauge sector contains twelve CP-even Wilson coefficients and an enlarged helicity-amplitude structure. Using a modified forward-limit dispersion relation, we analytically obtain non-trivial linear and non-linear elastic positivity constraints that define a spectrahedral geometry. We analyze implications of these bounds on non-forward amplitudes and discuss where kinetic-mixing and dark-axion-portal UV completions populate this geometry.
\end{abstract}

\maketitle

\vspace{0.2cm}
\noindent {\bf Introduction.}
%
Hidden-sector extensions of the Standard Model (SM) containing light, weakly coupled particles are well motivated theoretically and are central targets of current searches\,\cite{Essig:2013lka, Lanfranchi:2020crw}. Dark-photon extensions\,\cite{Fabbrichesi_2021, Caputo:2026pdw} are especially distinctive---they contain the only renormalizable, gauge-invariant portal built solely from gauge-field strengths that connects the SM to a hidden sector without assigning SM fields dark quantum numbers, namely the kinetic-mixing portal. They also have rich low-energy phenomenology\,\cite{Caputo:2026pdw,Jaeckel:2010ni} and are relevant to emerging frontiers\,\cite{Gori:2022vri}.

These considerations motivate a low-energy EFT of photons and a massless dark photon below the electron mass scale $m_e\sim\mathrm{MeV}$, assuming the dark-sector states are heavier. This photon--massless-dark-photon EFT is the appropriate theory whenever Mandelstam invariants, probe frequencies, and inverse background-variation scales satisfy $|s|,|t|,|u|,\omega^2,\partial^2\ll m_e^2$. This domain is the same one underlying the photon EFT\,\cite{2006physics5038H}, applicable for low-energy scatterings\,\cite{Karplus:1951zz,Milstein:1994zz}, Euler-Heisenberg light-by-light physics and low-energy probes of vacuum nonlinearity\,\cite{Battesti:2012hf,Marklund:2006my}, ranging from optical light-shining-through-wall and cavity experiments ($\omega_\gamma\sim \mathrm {eV}$)\,\cite{Jaeckel:2010ni,Redondo:2010dp}, to X-ray free electron laser vacuum-birefringence proposals ($\omega_{\mathrm{X}}\sim \mathrm {keV}$)\,\cite{Karbstein:2021bwi}, as well as astrophysical and cosmological settings with sub-electron-scale photons and electromagnetic fields ($\omega^{\mathrm{CMB}}\sim \mathrm{meV}$, $\omega^{\mathrm{NS}}_{\mathrm{X}}\sim \mathrm{keV}$)\,\cite{Harding:2006qn,Marklund:2006my,Hoseinpour:2020mfm}.

Low-energy EFTs are constrained by causality and S-matrix analyticity\,\cite{Adams:2006sv}, thereby restricting the space of viable infrared theories\,\cite{Vafa:2005ui,ArkaniHamed:2020blm}. Such consistency conditions have led to powerful bounds on low-energy physics\,\cite{deRham:2022hpx}. The photon--dark-photon EFT is qualitatively and quantitatively richer than the photon EFT\,\cite{Henriksson:2021ymi,Chowdhury:2021ynh,H_ring_2024} in this context---two massless spin-$1$ species, $\gamma$ and $\gamma'$, generate many more particle and helicity channels; crossing relates visible and dark external states; and the leading Wilson-coefficient space no longer reduces to the familiar two-parameter Euler-Heisenberg subspace\,\cite{2006physics5038H}. While gravitational positivity has been used to bound renormalizable parameters of specific dark-photon models\,\cite{Noumi:2022wwf, Aoki:2023khk, kim2025constrainingmillichargeddarkmatter}, and fixed-helicity multi-$U(1)$ positivity has been examined in a restricted sense in the tower weak-gravity setting\,\cite{Andriolo:2018lvp}, a general and detailed analysis of the photon--massless-dark-photon EFT and its positivity geometry remains, to our knowledge, unexplored.

In this paper, we derive positivity bounds on the dimension-$8$ Wilson coefficients of the photon--massless-dark-photon EFT. We enumerate the independent helicity amplitudes, finding $19$ independent amplitudes instead of the $3$ present in the pure-photon case, and construct the complete CP-even operator basis through dimension-$8$, obtaining $12$ Wilson coefficients instead of $2$. Using a modified forward-limit dispersion relation, we derive elastic positivity bounds that currently provide the strongest model-independent theoretical constraints on this EFT. In analytically tractable subspaces, we obtain explicit linear, quadratic, and nonlinear inequalities, several of which admit geometric interpretations as spectrahedral slices of the Wilson-coefficient cone or as shadows of a spectrahedral lift. These structures are in the same broad spirit as the larger EFT-hedron programme\,\cite{ArkaniHamed:2020blm}, where dispersive positivity organizes EFT data through convex hulls, cyclic polytopes, and Hankel total-positivity geometry. The bounds we obtain also imply hierarchies for several mixed-helicity amplitudes, as well as two-sided bounds for amplitudes, even in the non-forward limit. Finally, we show how kinetic-mixing and dark-axion-portal UV completions populate distinct regions of the resulting geometry, sharpening the map from IR Wilson coefficients to possible UV physics.

\vspace{0.2cm}
\noindent {\bf Independent Helicity Channels and Amplitudes.}
%
For $2\to 2$ scatterings involving photons ($\gamma$) and dark photons ($\gamma'$), in a $P$ and $CP$ symmetric theory, let us consider the number of independent helicity channels and the number of independent amplitude functions by systematically applying the parity-reversal and time-reversal transformation rules, along with the $s-t$, $s-u$ and $t-u$ crossing equations\,\cite{deRham:2017zjm, hebbar2022spinningsmatrixbootstrap4d}. As far as we are aware, a complete enumeration of the independent helicity channels and amplitude functions for mixed $\gamma-\gamma'$ scattering has not been explicitly analyzed in the literature. Helicity amplitudes will be denoted by $\mathcal{M}_{i,j}^{k,l}(s,t)$, where $(i,j)$ are the initial states and $(k,l)$ the final states, all indices running over both particle and helicity modes (i.e., $\{+, -, +', -'\}$). Where unambiguous, for conciseness, we will frequently drop the explicit $(s,t)$ arguments.

In $\gamma\gamma \to \gamma\gamma$ scattering, it is well-known\,\cite{Henriksson:2021ymi, H_ring_2024} that after using the transformation rules and crossing equations, one is left with $5$ independent helicity channels, which can be written in terms of $3$ independent amplitudes. For $\gamma'\gamma' \to \gamma'\gamma'$, the analysis and counting will be similar, and the independent helicity amplitudes may be associated with
\begin{equation}
\mathcal{M}_{+'+'}^{+'+'},~\mathcal{M}_{+'+'}^{-'-'},~\mathcal{M}_{+'+'}^{+'-'} \; .
\end{equation}
For the $\gamma\gamma' \to \gamma\gamma'$ particle channel, from parity and time-reversal transformation rules, there is a reduction in the number of helicity channels, from $16$ to $6$. Two of the helicity channel amplitudes may now be related by an $s-u$ crossing equation, giving $5$ independent amplitudes. These $5$ independent amplitudes may be associated with
\begin{equation}
\mathcal{M}_{++'}^{++'},~\mathcal{M}_{++'}^{+-'},~\mathcal{M}_{++'}^{-+'},~\mathcal{M}_{++'}^{--'},~\mathcal{M}_{+-'}^{-+'}  \; .
\end{equation}

In the $\gamma\gamma \to \gamma'\gamma'$ case, while parity transformation rules may be utilized in tandem with crossing equations to get a reduction to $6$ independent helicity channels, the time-reversal transformation relation does not lead to any further reductions. Using a $t-u$ crossing equation further gives $5$ independent helicity amplitudes. This counting is, of course, as expected from the earlier $\gamma\gamma' \to \gamma\gamma'$ analysis, based on their mutual crossing symmetry relations.

Coming to $\gamma\gamma \to \gamma\gamma'$, there is no reduction again in the independent helicity channels due to time-reversal transformation rules. Parity transformation rules give $8$ independent helicity channels. Crossing equations furnish four relations between these amplitudes, involving $s-u$ and $t-u$ swaps, finally giving $4$ independent helicity amplitudes that may be associated with
\begin{equation}
\mathcal{M}_{++}^{++'},~\mathcal{M}_{++}^{+-'},~\mathcal{M}_{++}^{-+'},~\mathcal{M}_{++}^{--'}  \; .
\end{equation}
The $\gamma'\gamma' \to \gamma'\gamma$ analysis and counting is similar to the above case, and the independent helicity amplitudes may be assigned to 
\begin{equation}
\mathcal{M}_{+'+'}^{+'+},~\mathcal{M}_{+'+'}^{+'-},~\mathcal{M}_{+'+'}^{-'+},~\mathcal{M}_{+'+'}^{-'-}  \; .
\end{equation}

Therefore, for $2\to 2$ scatterings involving massless photons and dark photons, analyzing all particle and helicity channels, one finds $19$ independent helicity amplitudes. This may be contrasted with the $3$ independent helicity amplitudes in $\gamma\gamma \to \gamma\gamma$ scattering\,\cite{Henriksson:2021ymi, H_ring_2024}.

\vspace{0.2cm}
\noindent {\bf Photon Dark-photon operators.}
%
Consider a low-energy EFT of massless photons and dark photons in $D=4$ flat spacetime. We will work in an operator basis where the kinetic terms are canonically normalized, meaning any kinetic-mixing term has already been eliminated by a field redefinition, and the physical photon and dark photon fields appear in the respective field tensors. 

We then find that there are $14$ CP-even EFT operators that are independent, up to dimension-$8$; in contrast to $3$ EFT operators in the photon EFT. There are the two kinetic terms for the photons and dark photons at dimension-$4$, no operators at dimension-$6$, and a total of $12$ operators at dimension-$8$. The latter are given by
\begin{equation}
    \label{operators}
    \begin{alignedat}{2}
        \mathcal{O}_1 &= (F\cdot F)^2,
        &\qquad \mathcal{O}_2 &= (F\cdot \tilde F)^2, \\
        \mathcal{O}_3 &= (F'\cdot F')^2,
        &\qquad \mathcal{O}_4 &= (F'\cdot \tilde F')^2, \\
        \mathcal{O}_5 &= (F\cdot F)(F'\cdot F'),
        &\qquad \mathcal{O}_6 &= (F\cdot \tilde F)(F'\cdot \tilde F'), \\
        \mathcal{O}_7 &= (F\cdot F')^2,
        &\qquad \mathcal{O}_8 &= (F\cdot \tilde F')^2, \\
        \mathcal{O}_9 &= (F\cdot F)(F\cdot F'),
        &\qquad \mathcal{O}_{10} &= (F\cdot \tilde F)(F\cdot \tilde F'), \\
        \mathcal{O}_{11} &= (F\cdot F')(F'\cdot F'),
        &\qquad \mathcal{O}_{12} &= (F\cdot \tilde F')(F'\cdot \tilde F') .
    \end{alignedat}
\end{equation}
Here, $F$ denotes $F_{\mu\nu}=\partial_\mu A_\nu-\partial_\nu A_\mu$, and $F'$ denotes $F'_{\mu\nu}=\partial_\mu A'_\nu-\partial_\nu A'_\mu$, with $A_\mu$ and $A'_\mu$ being the photon and the dark photon fields. The two leading-order Euler-Heisenberg operators\,\cite{2006physics5038H}, $\mathcal{O}_1$ and $\mathcal{O}_2$, are the only ones present in the photon EFT.

The EFT Lagrangian of interest, in terms of these operators, is
\begin{equation}
    \label{le-lagrangian}
    \mathcal{L}_{\gamma\gamma'}^{\mathrm{EFT}}
    = -\frac{1}{4} F_{\mu\nu}F^{\mu\nu}
    -\frac{1}{4} F'_{\mu\nu}F'^{\mu\nu}
    + \sum_{i=1}^{12} c_i\,\mathcal{O}_i+\ldots \; .
\end{equation}
$c_i$ are the dimension-8 Wilson coefficients and $\ldots$ represents the next-to-leading order EFT operators, above dimension-$8$. Our aim is to place model-agnostic, general bounds from unitarity on the low-energy Wilson coefficients $c_i$. Towards this, we systematically consider the contribution of the above operators to the various $2\to 2$ particle and helicity channel scattering amplitudes---for example, $\mathcal{M}_{+,+'}^{-,-'}= 4t^2 (c_5 - c_6) + 2(s^2 + u^2) (c_7 - c_8) $, $\mathcal{M}_{+,+}^{-,-'}= 2(s^2 + t^2 + u^2) (c_9 - c_{10})$, $\mathcal{M}_{+',-'}^{+,-'} = 2u^2 (c_{11} + c_{12})$. The helicity amplitudes have a characteristic dependence on $c_k \pm c_{k+1} $, which may be traced to an alternative representation of Eq.\,\eqref{le-lagrangian} in terms of Hodge self-dual and anti-self-dual operators, which correspond to fixed-helicity states.

\vspace{0.2cm}
\noindent {\bf Elastic Positivity and Wilson Coefficient Constraints.}
%
Express an arbitrary state $\ket{A}$ as $\ket{A}=\sum_{i,j} a_{i}b_{j} \ket{i,j}$, with $\ket{i,j}\equiv \ket{i}\otimes \ket{j}$ and $i, j\in \{+, -, +', -'\}$. The corresponding scattering amplitude for $A\to A$ can be written as $\mathcal{M}(s)=\sum_{i,j,k,l} \mathcal{M}_{i,j}^{k,l} a_{i}b_{j} a^{*}_{k}b^*_{l}$.

We use a modified dispersion integral construction compared to pure photon EFT\,\cite{Henriksson:2021ymi, H_ring_2024}, and generalize other constructions where fixed helicities had to be assumed\,\cite{Andriolo:2018lvp} to implement $s-u$ symmetry and where an incomplete branch cut was adopted. Towards this, define in the open upper-half $s$-plane, the analytic function
\begin{equation}
\mathscr{M}(s)=\frac{\mathcal{M}(s)+\mathcal{M}(-s)}{s^3}\;.
\end{equation}
This is, by construction, odd under $s-u$ crossing in the forward limit under consideration. Mindful of the branch cut across the whole real $s$-axis, in the massless photon--dark-photon EFT, consider a typical closed contour\,\cite{Bellazzini:2020cot} in the upper-half plane, where $\mathscr{M}(s)$ is analytic. Obstructions due to graviton exchange are neglected by taking the strict decoupling limit $M_{\mathrm{Pl}}/\Lambda_{\mathrm{EFT}} \to \infty$\,\cite{Alberte:2020jsk}, or by assuming Reggeized gravitational exchange with only parametrically suppressed residual corrections\,\cite{Tokuda:2020mlf}.

Now, from the crossing equations\,\cite{deRham:2017zjm,hebbar2022spinningsmatrixbootstrap4d}, one can show (see the End Matter) that $\mathcal{M}(-s-i\epsilon)=\sum_{i,j,k,l} \mathcal{M}_{i,j}^{k,l} (s+i\epsilon)\,\tilde{a}_{i}b_{j} \tilde{a}^{*}_{k}b^*_{l}$, with $\tilde{a}_i=a^*_{-i}$. This may be interpreted as the scattering amplitude corresponding to transfer matrix element $\bra{\tilde{A}}\hat{T}\ket{\tilde{A}}$, with $\ket{\tilde{A}}=\sum_{i,j}\tilde{a}_i b_j\ket{i,j}$. Using the optical theorem, one then concludes that
\begin{equation}\label{eq:ImMminuss}
\Im\mathcal{M}_{A\to A}(-s-i\epsilon)=\Im\mathcal{M}_{\tilde{A} \to \tilde{A} }(s+i\epsilon)\geq0 \; .
\end{equation}
Combining the inferences above with the assumption of Regge boundedness for massless spin-1 amplitudes\,\cite{Chowdhury:2019kaq, Chandorkar:2021viw} leads to the result 
\begin{equation}
    \int_{\mathcal{C}_0} \frac{ds}{2\pi i}\,\mathscr{M}(s) \geq 0 \; .
\end{equation}
Here, $\mathcal{C}_0$ is the arc in the infrared circumscribing $s=0$. In terms of the helicity amplitudes $\mathcal{M}_{i,j}^{k,l}(s)$ in the forward limit, this gives, after using the $s-u$ crossing equations, the elastic positivity condition
\begin{equation}\label{eq:blockpositivity}
\sum_{i,j,k,l} a_{i}b_{j} a^*_{k}b^*_{l}
        \int_{\mathcal{C}_0} \frac{ds}{2\pi i}\,
        \frac{\mathcal{M}_{i,j}^{k,l}(s)
        +\mathcal{M}_{-k,j}^{-i,l}(s)}{s^3}
        \geq 0 \;.
\end{equation}
The amplitude integral above may be interpreted as a convex cone, and $(a_i b_j a_k^* b_l^*)$ are in the dual convex cone\,\cite{Zhang:2020jyn, Li:2021lpe}. While the full dual convex cone encodes more information\,\cite{Zhang:2020jyn, Li:2021lpe} in principle than Eq.\,\eqref{eq:blockpositivity}, we will nevertheless be able to derive several non-trivial slices of the full positivity cone, yielding the sharpest bounds presently known. Eq.\,\eqref{eq:blockpositivity} may also be given an algebraic-geometric interpretation as positivity on the Segre variety $\operatorname{Seg}(\mathbb P^3\times\mathbb P^3)\subset\mathbb P^{15}$, the projective variety of rank-one tensors corresponding to factorized two-particle states, formed from $\{+, -, +', -'\}$.

We now apply the elastic positivity condition to explicitly derive several non-trivial bounds on the dimension-8 Wilson coefficients in Eq.\,\eqref{le-lagrangian}. The full $a \otimes b \in \C^4 \otimes \C^4$ elastic positivity may be studied systematically in principle, either analytically or by semidefinite programming techniques\,\cite{Vandenberghe:1996SDP}. For brevity, we will present explicit results from some analytically tractable $\C^2 \otimes \C^4$ subspaces here. A subset of our $\C^2 \otimes \C^2$ subspace bounds is implicit in\,\cite{Andriolo:2018lvp}, which investigated the tower weak gravity conjecture. We also point out that some of the non-linear constraints we derive below succinctly encode an infinitude of linear inequalities.
 
For conciseness of expressions, define $S_k=c_k+c_{k+1}$ and $D_k=c_k-c_{k+1}$. Analysis of elastic positivity in some of the $\C^2 \otimes \C^4$ subspaces gives the following compact, independent constraints
\begin{gather}
    \label{linear}
    c_1, c_2, c_3, c_4, c_7, c_8 \ge 0 \; , \\
    \label{quadratic-1}
    4 c_1 c_7 \ge c_9^2 \, , \qquad 4 c_2 c_8 \ge c_{10}^2 \; , \\
    \label{quadratic-2}
    4 c_3 c_7 \ge c_{11}^2 \, , \qquad 4 c_4 c_8 \ge c_{12}^2 \; , \\
    \label{Sylvester1}
    4 S_1 S_3 \ge S_5^2 \; , \\
    \label{elliptope}
    4 S_1 S_3 S_7 + S_5 S_9 S_{11} \ge S_7 S_5^2 + S_1 S_{11}^2 + S_3 S_9^2 \; , \\
    \label{quarticinvariant2}
    C_2 \ge -2\sqrt{C_0 C_4} \; ,\\
    \label{quarticinvariant3}
    4 C_0 C_2 - C_1^2 \ge -8 C_0 \sqrt{C_0 C_4} \; ,\\
    \label{quarticinvariant1}
    4 H_2^3 - H_3^2 \ge 0 \; ,\\
    \label{quarticinvariant4}
    H_1 \ge 0 \qquad \text{or} \qquad 16 C_4^2 H_2 - H_1^2 \ge 0\; .
\end{gather}
Here, $C_k$ are the coefficients of a positive-semidefinite quartic polynomial, defined in terms of $S_k$ and $D_k$, and $H_k$ are its quartic invariants (defined in the End Matter). The above general, independent bounds may now be resolved and parsed by considering a few prominent subspaces of $\C^2 \otimes \C^4$ as special cases. 

\begin{figure}[t]
\centering
\includegraphics[width=\columnwidth]{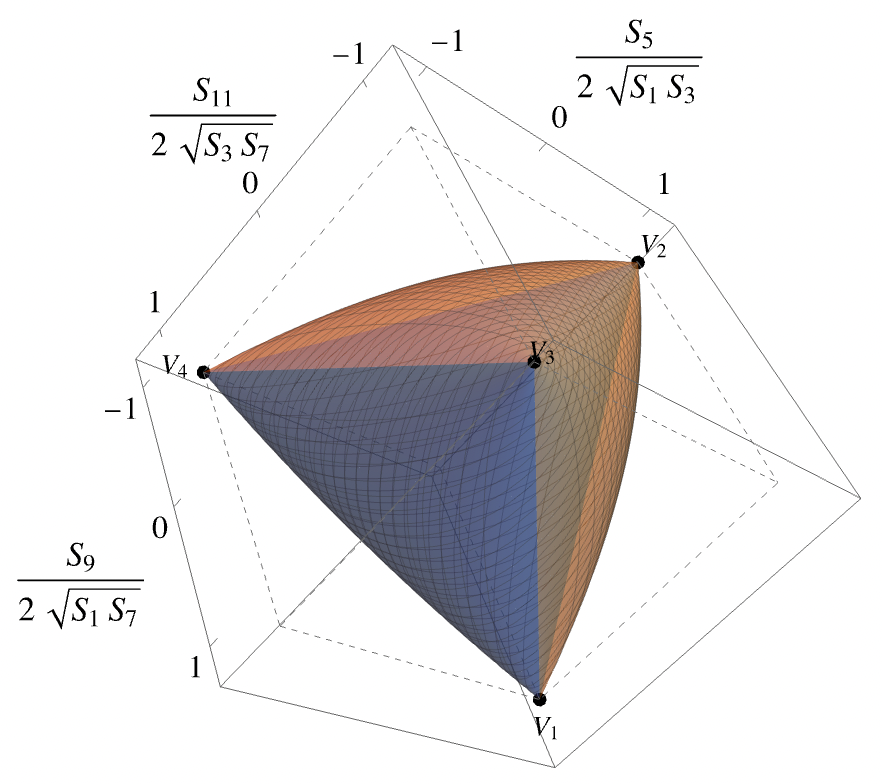}
\caption{Spectrahedron defining an allowed three-dimensional positivity region, for a specific set of three Wilson coefficient combinations $(\frac{S_5}{2 \sqrt{S_1 S_3}}, \frac{S_9}{2 \sqrt{S_1 S_7}}, \frac{S_{11}}{2 \sqrt{S_3 S_7}})$, taken as axes. The vertices $(V_1,V_2,V_3,V_4)$ are located at $(1,1,1),\,(1,-1,-1),\,(-1,1,-1)$, and $(-1,-1,1)$, respectively. The surface of the elliptope, given by $\mathbb{V}^{\,2}_{\smallparallelepiped} = 0$, is Cayley's four-nodal cubic surface\,\cite{laurent2026algebraicboundarygraphelliptopes} and the elliptope volume is $\mathbb{V}(\mathcal{E}^3)=\pi^2/2$.}
\label{fig:three_coefficient_elliptope}
\end{figure}

Consider first, elastic positivity in some of the $\C^2 \otimes \C^2$ subspaces. Then, one of the inferences from Eqs.\,\eqref{linear}-\eqref{elliptope} is that 
\begin{equation}\label{eq:Gram1}
G_S=\left(\begin{array}{ccc}
4 S_1 & S_9 & 2 S_5 \\
S_9 & S_7 & S_{11} \\
2 S_5 & S_{11} & 4 S_3
\end{array}\right) \succeq 0 \; .
\end{equation}
In cases where elastic positivity reduces to a positive semidefiniteness criterion, as in Eq.\,\eqref{eq:Gram1}, we may give an interpretation in terms of Gram matrices and posit a geometric interpretation of the bounds as correlation matrix spectrahedra\footnote{A spectrahedron is a convex shape defined by a linear matrix inequality $M_0+p_1 M_1+p_2 M_2+\cdots+p_k M_k \succeq 0$, with real $p_i$, and fixed symmetric matrices $M_i$. The $p_i$ correspond to the Wilson coefficients $c_i$ in our case.}, referred to as elliptopes. An elliptope demarcating an admissible three-dimensional region, for the Wilson coefficient combinations $(\frac{S_5}{2 \sqrt{S_1 S_3}}, \frac{S_9}{2 \sqrt{S_1 S_7}}, \frac{S_{11}}{2 \sqrt{S_3 S_7}})$, is shown in Fig.\,\ref{fig:three_coefficient_elliptope}. The boundary corresponds to the determinant-zero locus and the four sharp vertices correspond to rank-$1$ configurations. For visual separation, the blue and orange shadings on the convex boundary denote two branches of the determinant-zero solution for $\frac{S_{11}}{2 \sqrt{S_3 S_7}}$. In this context, just the positivity criterion in Eq.\,\eqref{elliptope} admits as well a volume interpretation, in terms of the Euclidean volume of the parallelepiped formed by the Gram vectors, $\det (G_S)\equiv \mathbb{V}^{\,2}_{\smallparallelepiped} \geq 0$.

From this subspace, one is also able to extract constraints of the form
\begin{gather}
\label{Sylvester2}
4 S_1 S_7 \geq D_9^2,\quad 4 S_3 S_7 \geq D_{11}^2 \; , \\
|2D_5+D_7| \leq S_7+4 \sqrt{S_1 S_3} \; . 
\end{gather}

Eqs.\,\eqref{quarticinvariant2}-\,\eqref{quarticinvariant4} are equivalent to the matrix polynomial positive semidefiniteness criterion
\begin{equation}\label{eq:spectrahedronshadow}
P_4(c_i,x) \equiv x^2 T_2+x T_1+T_0 \succeq 0, \quad \forall x \in \mathbb{R} \; ,
\end{equation}
where $T_2,T_1,T_0$ are $2\times2$ symmetric matrices of $S_k$ and $D_k$. Eq.\,\eqref{eq:spectrahedronshadow}, defines an infinite intersection of spectrahedra. In general, an infinite intersection of spectrahedra need not itself be a spectrahedron, but in the present case, by the S-procedure\,\cite{PolikSProcedure}, it admits a spectrahedral lift\,\cite{GouveiaParriloThomas2013}, defined by
\begin{equation}
P_4(c_i,x) \succeq 0 ~\forall x \in \mathbb{R} ~ \Longleftrightarrow ~ \exists \lambda \in \mathbb{R} \mid \mathcal{P}_4(c_i,\lambda) \succeq 0 \;.
\end{equation}
This lifted set defined by $\mathcal{P}_4(c_i,\lambda)\succeq 0$ is a genuine spectrahedron, as it can be expressed as a linear matrix inequality in the Wilson coefficients $c_i$ and lift variable $\lambda$, and can be computed analytically (see End Matter). Therefore, the condition $ P_4(c_i,x) \succeq 0 ~ \forall x \in \mathbb{R}$ defines a spectrahedral shadow\,\cite{Scheiderer2018}  in the Wilson coefficient space---i.e. a projection of the higher-dimensional spectrahedron defined by $\mathcal{P}_4(c_i,\lambda)\succeq 0$. This is the geometric interpretation of the quartic constraints in Eqs.\,\eqref{quarticinvariant2}-\eqref{quarticinvariant4}.

The Wilson coefficient bounds contained in Eqs.\,\eqref{linear}-\,\eqref{quarticinvariant4} are rich in physical content. We now examine several analytically tractable consequences.

Consider Eq.\,\eqref{linear}. Unlike $c_{1,2}\geq 0$ in the photon EFT case, the additional positivity constraints $c_3, c_4, c_7, c_8 \ge 0$ are potentially meaningful phenomenological restrictions at low-energy. This is because, it is now strongly supported by the absence of significant deviations in $(g-2)_\mu$\,\cite{Aguillard:2025mug2final,Aliberti:2025g2update}---to which SM Delbr\"{u}ck-like and Hadronic light-by-light scatterings\,\cite{Aldins:1970id,Kinoshita:2004wi,Kurz:2015bia} contribute---that $c_{1,2}$ are dominated by QED. Specifically, $c_{1,2} \simeq \alpha^2/(90 \,m_e^4)\left(1+\delta c_{1,2}^{\text{\tiny{SM}}}+\delta c_{1,2}^{\text{\tiny{NP}}}\right)$, with $\delta c_{1,2}^{\text{\tiny{SM/NP}}} \ll 1$ denoting all sub-leading contributions from the SM and any unknown new-physics. The photon EFT positivity bounds for these Wilson coefficients are then just implying the trivial condition $\left(1+\delta c_{1,2}^{\text{\tiny{SM}}}+\delta c_{1,2}^{\text{\tiny{NP}}}\right) \geq 0$. In contrast, there is a priori no stipulation that $c_i \,\forall i >2$ should be SM dominated. 

Next, note that Eqs.\,\eqref{linear}-\eqref{quadratic-2} also imply that $4 S_1 S_7 \geq S_9^2$ and $4 S_3 S_7 \geq S_{11}^2$. Along with Eq.\,\eqref{Sylvester1}, these translate to a hierarchy of low-energy amplitudes in the physical region
\begin{equation}
    \label{eq:amplhierarchy}
    \begin{aligned}
        \left| \mathcal{M}_{+'+'}^{++}(s,t) \right|^2 &\lesssim \widebar{\mathcal{M}}_{++}^{++}(s,t) ~\widebar{\mathcal{M}}_{+'+'}^{+'+'}(s,t) \; ,\\
        \left| \mathcal{M}_{++}^{++'}(s,t) \right|^2 &\lesssim \widebar{\mathcal{M}}_{++}^{++}(s,t) ~\widebar{\mathcal{M}}_{++'}^{++'}(s,t) \;, \\
        \left| \mathcal{M}_{+'+'}^{+'+}(s,t) \right|^2 &\lesssim \widebar{\mathcal{M}}_{+'+'}^{+'+'}(s,t) ~\widebar{\mathcal{M}}_{++'}^{++'}(s,t) \; .
    \end{aligned}
\end{equation}
Note that this hierarchy is valid even in the non-forward limit ($t\neq 0$). Here, and in below, we define $\widebar{\mathcal{M}}_{i,j}^{k,l}=\Re\mathcal{M}_{i,j}^{k,l}$.
Also, as a corollary, we deduce that to leading order if in some kinematic region $\mathcal{M}_{++'}^{++'}(s,t)\to 0$ it necessarily implies that $\mathcal{M}_{++}^{++'}(s,t) \to 0$ and $\mathcal{M}_{+'+'}^{+'+}(s,t) \to 0$ as well. 

Additionally, Eq.\,\eqref{elliptope} implies two-sided bounds on various low-energy amplitudes, credible even in the non-forward limit. For instance, for $S_7 > 0$ in Eq.\,\eqref{linear}, we have the two-sided bound on $\widebar{\mathcal{M}}_{+' +'}^{++}(s,t)$ in the physical region
\begin{equation}
\Delta_-(s,t) \lesssim \widebar{\mathcal{M}}_{+' +'}^{++}(s,t) \lesssim \Delta_+(s,t) \; .
\end{equation}
The closed interval is circumscribed by a function of the other helicity amplitudes
\begin{multline}
\Delta_\pm = \frac{\widebar{\mathcal{M}}_{++}^{++'} \widebar{\mathcal{M}}_{+'+'}^{+'+}}{\widebar{\mathcal{M}}_{++'}^{++'}}
\pm \left[\widebar{\mathcal{M}}_{++}^{++}-\frac{|\mathcal{M}_{++}^{++'}|^2}{\widebar{\mathcal{M}}_{++'}^{++'}}\right]^{\frac{1}{2}} \\
\times\left[\widebar{\mathcal{M}}_{+'+'}^{+'+'}-\frac{|\mathcal{M}_{+'+'}^{+'+}|^2}{\widebar{\mathcal{M}}_{++'}^{++'}}\right]^{\frac{1}{2}} .
\end{multline}

Now, consider the case of a kinetic mixing portal\,\cite{HOLDOM1986196, Galison:1983pa} UV completion with two $U(1)_{a,b}$ groups, a SM fermion ($e$), a dark fermion ($e'$), gauge couplings ($g_{a,b}$), kinetic mixing parameter $\varepsilon$ and an $SO(2)$ angle $\theta$ characterizing relative couplings\,\cite{Fabbrichesi_2021}. In this UV completion, integrating out the Dirac fermions at 1-loop order gives the leading-order contributions to $c_i^{\mathrm{K}}(g_{a,b},m_{e,e'},\varepsilon,\theta)$---for instance, $2c_5^{\mathrm{K}} = \frac{8}{7}c_6^{\mathrm{K}} = c_7^{\mathrm{K}} = \frac{4}{7}c_8^{\mathrm{K}} = \frac{1}{(1-\varepsilon^2)^2} \bigl[\frac{\alpha_a^2}{90m_e^4} \sin^2(2\theta)+ \frac{\alpha_b^2}{90m_{e'}^{4}}  \sin^2(2\theta-2\sin^{-1}\varepsilon)\bigr]$. The  $c_i^{\mathrm{K}}$ are found to satisfy the bounds, but which bounds are saturated crucially depends on the details of the kinetic mixing portal UV completion.

When $\theta=0$, implying UV completions where $e$ does not interact with $\gamma'$ and $e'$ is fractionally charged with respect to $\gamma$ (i.e. ``millicharged'' particle models), we find that the bounds in Eqs.\,\eqref{quadratic-2}, \eqref{elliptope}, \eqref{quarticinvariant3}, and \eqref{quarticinvariant1} are saturated.
Moreover, since $S_{11} = -2\sqrt{S_3 S_7}$ in this case, the UV completions all lie on an edge of the IR elliptope in Fig.\,\ref{fig:three_coefficient_elliptope}. Specifically, it lies on the line $\left(V_2, \frac{V_2+V_3}{2}\right]$. In contrast, for $\theta=\sin^{-1}\varepsilon$, defining UV completions where $e$ interacts with $\gamma'$ but $e'$ does not couple to $\gamma$, the bounds in Eqs.\,\eqref{quadratic-1}, \eqref{elliptope}, \eqref{quarticinvariant1} and \eqref{quarticinvariant4} are saturated.
In this case, as $S_9 = -2\sqrt{S_1 S_7}$, all the UV completions lie on line $\left(V_2, \frac{V_2+V_4}{2}\right]$. In general, kinetic mixing UV portals with two fermions furnish rank-$2$ contributions to $G_S$ of Eq.\,\eqref{eq:Gram1} and thereby lie on the surface of the IR elliptope.
In kinetic mixing portal UV completions where there are effectively no dark fermions ($\frac{m_{e'}}{m_e}\to \infty$), we find that Eqs.\,\eqref{quadratic-1}-\eqref{quarticinvariant4} are saturated. This implies that all kinetic mixing portal UV completions with no matter fields in the dark sector must lie on a vertex of the IR elliptope in Fig.\,\ref{fig:three_coefficient_elliptope}, specifically $V_2$ for $0\le\theta<\pi/2$ and $V_1$ for $\pi/2<\theta\le\pi$. This may also be understood from the fact that in this case, the contributions to $G_S$ are of rank $1$.

In an effective tree-level UV completion like the (dark) axion portal\,\cite{Peccei:1977hh, Kaneta:2016wvf}, with a massive pseudoscalar ($\varphi$) coupling to $\gamma$ and $\gamma'$, with dark axion couplings $g$, $g'$ and $g_\varepsilon$, one obtains after integrating out the dark axion at tree-level, contributions to the dimension-8 Wilson coefficients $c_2^{\mathrm{A}}$, $c_4^{\mathrm{A}}$, $c_6^{\mathrm{A}}$, $c_8^{\mathrm{A}}$, $c_{10}^{\mathrm{A}}$ and $c_{12}^{\mathrm{A}}$. These saturate the bounds in Eqs.\,\eqref{quadratic-1}-\eqref{quarticinvariant4}. It is found that the UV completion lies on the vertices of Fig.\,\ref{fig:three_coefficient_elliptope}, specifically as $\left(\sgn(g  g'),\sgn(g g_\varepsilon),\sgn(g' g_\varepsilon)\right) \in V_i $. Finally, if the kinetic mixing portal and dark axion portal concurrently exist in the UV, it is seen that such a hybrid UV completion would then exist inside the elliptope in the IR. This is also generally true when the UV sector is richer. This is because in these cases, the contributions are generically of rank $3$. Additional IR positivity structures of the photon--massless-dark-photon EFT may yield further refinements in these cases.

In summary, this paper derives a non-trivial positivity geometry for the photon--massless-dark-photon EFT, connecting it to model-independent hierarchies, two-sided constraints on mixed photon--dark-photon amplitudes, and well-motivated UV completions. Our results constitute the first complete positivity analysis in this wide and currently relevant setting and provide the strongest bounds presently available, while furnishing sharp IR diagnostics of viable dark-sector UV physics.


\medskip
\noindent {\it Acknowledgments.}
SC is supported by a KVPY fellowship, and YD by an INSPIRE fellowship.
SC and YD also gratefully acknowledge financial support from the Infosys and IDeaS foundations.
 Parts of this work were completed during the ``Amplitudes 2026'' conference at Queen Mary University of London. AT thanks the organizers and acknowledges useful discussions during the meeting. We acknowledge the use of \texttt{Wolfram Mathematica}\,\cite{Mathematica} for a few of the symbolic computations. \texttt{FeynRules}\,\cite{Christensen_2009,Christensen_2011,Alloul_2014}, \texttt{FeynCalc}\,\cite{Shtabovenko_2016,Shtabovenko_2020,Shtabovenko_2025,MERTIG1991345}, \texttt{FeynArts}\,\cite{Hahn_2001}, and \texttt{Package-X}\,\cite{Patel_2015,Patel_2017} were used for some of the UV model computations. AI-assisted tools were used only for typographical and grammatical proofreading of author-written text.
\bibliography{DarkPhotonEFT}

\appendix
\onecolumngrid
\section*{End Matter}
\twocolumngrid
\setcounter{equation}{0}

\renewcommand{\theequation}{E.\arabic{equation}}
\renewcommand{\theHequation}{EndMatter.\arabic{equation}}

\vspace{0.2cm}
\noindent {\bf Positive semidefiniteness of $\Im\mathcal{M}_{A\to A}(-s-i\epsilon)$}

We have
\begin{equation}
    \begin{split}
        \mathcal{M}(-s-i \epsilon)&=\sum_{i,j,k,l} \mathcal{M}_{i,j}^{k,l} (-s-i \epsilon)a_{i}b_{j} a^{*}_{k}b^*_{l}\\
        &=\sum_{i,j,k,l} \mathcal{M}_{-k,j}^{-i,l}(s+i \epsilon) a_{i}b_{j} a^{*}_{k}b^*_{l}\\
        &=\sum_{i,j,k,l} \mathcal{M}_{i,j}^{k,l} (s+i \epsilon)a_{-k}b_{j} a^{*}_{-i}b^*_{l}\;,\\
    \end{split}
\end{equation}
using the s-u crossing equation\,\cite{hebbar2022spinningsmatrixbootstrap4d}
\begin{align}
    {\mathcal{M}_{1234}}_{\lambda_1,\lambda_2}^{\lambda_3,\lambda_4}
    (k_1,k_2,k_3,k_4)
    &= \zeta_{13}
    {\mathcal{M}_{\bar{3}2\bar{1}4}}_{-\lambda_3,\lambda_2}^{-\lambda_1,\lambda_4}
    \notag\\
    &\quad \times (-k_3,k_2,-k_1,k_4) \;.
\end{align}
It can be shown that the crossing phase in our case $\zeta_{\gamma\gamma'} = 1$, in the forward limit\,\cite{deRham:2017zjm}. Defining $\tilde{a}_i=a^*_{-i}$, and re-indexing, gives
\begin{equation}
    \mathcal{M}(-s-i \epsilon)=\sum_{i,j,k,l} \mathcal{M}_{i,j}^{k,l} (s+i \epsilon)\tilde{a}_{i}b_{j} \tilde{a}^{*}_{k}b^*_{l} \; .
\end{equation}
This is the invariant matrix element associated with $\bra{\tilde{A}}\hat{T}\ket{\tilde{A}}$, for $\ket{\tilde{A}}=\sum_{i,j}\tilde{a}_i b_j\ket{i,j}$, and optical theorem should apply, thereby giving
\begin{equation}
    \Im\mathcal{M}_{A\to A}(-s-i \epsilon)=\Im\mathcal{M}_{\tilde{A}\to \tilde{A}}(s+i \epsilon)\geq0 \; .
\end{equation}

\vspace{0.2cm}
\noindent {\bf Functions of Wilson Coefficients and Quartic Invariants}

We define
\begin{gather}
    C_0 = 4 S_3 S_7 - S_{11}^2 \; ,\\
    C_1 = 8 S_3 D_9 + 2 S_7 D_{11} - 2 (2D_5+D_7) S_{11} \; ,\\
    C_2 = 16 S_1 S_3 + 4 D_9 D_{11} + S_7^2 - (2D_5+D_7)^2 - 2 S_9 S_{11} \; ,\\
    C_3 = 8 S_1 D_{11} + 2 S_7 D_9 - 2 (2D_5+D_7) S_9 \; ,\\
    C_4 = 4 S_1 S_7 - S_9^2 \; ,
\end{gather}
and based on these, the quartic invariants
\begin{gather}
H_1 = 8 C_2 C_4 - 3 C_3^2 \; ,\\
H_2 = C_2^2 - 3 C_1 C_3 + 12 C_0 C_4 \; ,\\
H_3 = 2 C_2^3 - 9 C_1 C_2 C_3 + 27 C_1^2 C_4 + 27 C_3^2 C_0 - 72 C_0 C_2 C_4
\end{gather}
\vspace{0.2cm}
\noindent {\bf Spectrahedral Lift and Shadow}
\begin{gather}
\begin{aligned}
T_2 & =\left(\begin{array}{cc}
4 S_1 & S_9 \\
S_9 & S_7
\end{array}\right), \quad T_0=\left(\begin{array}{cc}
S_7 & S_{11} \\
S_{11} & 4 S_3
\end{array}\right) \; , \\
T_1 & =\left(\begin{array}{cc}
2 D_9 & 2 D_5+D_7 \\
2 D_5+D_7 & 2 D_{11}
\end{array}\right) \; .
\end{aligned} \\
\mathcal{P}_4(\lambda)=\left(\begin{array}{cc}
T_0 & \frac{1}{2} T_1 +\lambda J \\
\frac{1}{2} T_1-\lambda J & T_2
\end{array}\right),~ J=\left(\begin{array}{cc}
0 & 1 \\
-1 & 0
\end{array}\right) \; .
\end{gather}

\vspace{0.2cm}
\noindent {\bf Kinetic Mixing and Dark Axion Portal Conventions}

Kinetic mixing portal UV completion is defined by
\begin{equation}
    \label{dark-photon-lagrangian}
    \begin{multlined}
        \mathcal{L}^{\mathrm{K}}
        \supset -\frac{1}{4}F^a_{\mu\nu}F_a^{\mu\nu}
            -\frac{1}{4}F^b_{\mu\nu}F_b^{\mu\nu}
            -\frac{\varepsilon}{2}F^a_{\mu\nu}F_b^{\mu\nu}
        \\
        + 
            e_a J^{e}_\mu A_a^\mu + e_b J^{e'}_\mu A_b^\mu
        \; .
    \end{multlined}
\end{equation}
The kinetic terms may be canonically normalised through
\begin{equation}
    \label{diagonalisation}
    \begin{pmatrix} A^a_\mu \\ A^b_\mu
    \end{pmatrix}
    =
    \begin{pmatrix}
        (1-\varepsilon^2)^{-1/2} & 0 \\
        -\varepsilon(1-\varepsilon^2)^{-1/2} & 1
    \end{pmatrix}
    \begin{pmatrix}
        \cos\theta & -\sin\theta \\
        \sin\theta & \cos\theta
    \end{pmatrix}
    \begin{pmatrix} A_\mu \\ A'_\mu
    \end{pmatrix} .
\end{equation}
Integrating out the SM fermion $e$ and the dark fermion $e'$, at 1-loop, gives, to leading order in gauge couplings, the low-energy dimension-8 Wilson coefficients
\begin{align}
c_1^{\mathrm{K}}&=\frac{4}{7}c_2^{\mathrm{K}}= \sec^4{\phi} \bigl[\beta_a\cos^4\theta +\beta_b\sin^4(\theta-\phi)\bigr], \\
c_3^{\mathrm{K}}&=\frac{4}{7}c_4^{\mathrm{K}} = \sec^4{\phi} \bigl[\beta_a\sin^4\theta +\beta_b\cos^4(\theta-\phi)\bigr] ,\\
\notag 2c_5^{\mathrm{K}}&=\frac{8}{7}c_6^{\mathrm{K}}=c_7^{\mathrm{K}}=\frac{4}{7}c_8^{\mathrm{K}} \\
&= \sec^4{\phi} \bigl[\beta_a\sin^2(2\theta)+\beta_b\sin^2(2\theta-2\phi)\bigr] ,\\
\notag c_9^{\mathrm{K}}&=\frac{4}{7}c_{10}^{\mathrm{K}}= 4\sec^4{\phi}\bigl[-\beta_a\sin\theta\cos^3\theta \\
&~~~~~~~~~~+\beta_b\sin^3(\theta-\phi)\cos(\theta-\phi)\bigr],\\
\notag c_{11}^{\mathrm{K}}&=\frac{4}{7}c_{12}^{\mathrm{K}}= 4\sec^4{\phi} \bigl[-\beta_a\sin^3\theta\cos\theta\\
&~~~~~~~~~~+\beta_b\sin(\theta-\phi)\cos^3(\theta-\phi)\bigr] .
\end{align}
We have defined $\beta_i=\frac{\alpha_i^2}{90m_i^4},\,i\in \{a,b\}$, with the identification $m_a \equiv m_e, m_b \equiv m_{e'}$, and $\phi=\sin^{-1}\varepsilon$.

Our dark axion portal UV completion is defined by the effective Lagrangian
\begin{equation}
\mathcal{L}^{\mathrm{A}} \supset \frac{\varphi}{4}\left(g F_{\mu \nu} \tilde{F}^{\mu \nu}+g' F_{\mu \nu}^{\prime} \tilde{F}^{\prime \mu \nu}+g_\varepsilon F_{\mu \nu}^{\prime} \tilde{F}^{\mu \nu} \right) \; .
\end{equation}
Integrating out the dark axion at tree level gives the leading order contributions to the dimension-8 Wilson coefficients. The only non-zero low-energy dimension-8 Wilson coefficients, at this order in the dark axion couplings, are
\begin{equation}
    \begin{aligned}
c_2^{\mathrm{A}} &=\frac{g^2}{32 m_\varphi^2}, \quad c_4^{\mathrm{A}}=\frac{g'{}^2}{32 m_\varphi^2}, \quad c_6^{\mathrm{A}}=\frac{g g'}{16 m_\varphi^2} \; , \\
c_8^{\mathrm{A}} &= \frac{g_\varepsilon^2}{32 m_\varphi^2}, \quad c_{10}^{\mathrm{A}} = \frac{g g_\varepsilon}{16 m_\varphi^2}, \quad c_{12}^{\mathrm{A}} = \frac{g' g_\varepsilon}{16 m_\varphi^2} \; .
\end{aligned}
\end{equation}


\end{document}